\documentclass[epjST]{svjour}
\usepackage{graphics}
\usepackage{bm}
\begin{document}
\title{Exact spin-orbit qubit manipulation}
\author{ Anton Ram\v{s}ak\inst{1,2}\fnmsep\thanks{\email{anton.ramsak@fmf.uni-lj.si}} \and Tilen {\v C}ade{\v z}\inst{3} \and 
Ambro{\v z} Kregar\inst{4,5} \and Lara Ul{\v c}akar\inst{2}  }
\institute{Faculty of Mathematics and Physics, University of Ljubljana, Ljubljana, Slovenia \and Jo{\v z}ef Stefan Institute, Ljubljana, Slovenia \and Beijing Computational Science Research Center, Beijing 100193, China \and Faculty of Education, University of Ljubljana, Ljubljana, Slovenia \and Faculty of Mechanical Engineering, University of Ljubljana, Ljubljana, Slovenia}

\abstract{
We consider exactly solvable  manipulation of spin-qubits  confined in a moving harmonic trap and in the presence of the time dependent Rashba interaction. Non-adiabatic Anandan phase for cyclic time evolution is compared to the Wilczek-Zee adiabatic counterpart. It is shown that the ratio of these two phases can for a chosen system
be any real number. Next we demonstrate the possibility of arbitrary qubit transformation in a ring with spin-orbit interaction. Finally, we present an example of exact analysis of spin-orbit dynamics influenced by the Ornstein-Uhlenbeck coloured noise.
} 
\maketitle
\section{Introduction}
\label{intro}
Spintronics as the new branch of electronics  has the potential for realising building blocks of a quantum computer via electron spin qubits. Implementation of such qubits is relatively simple in gated semiconductor devices based on quantum dots and quantum wires \cite{wolf01,hanson07}. Qubit manipulation may be achieved through rotation of the electron's spin  by the application of an external magnetic field \cite{Koopens06} and   by methods where magnetic field is not needed due to the use of the spin-orbit interaction (SOI)  \cite{dresselhaus55,bychkov84}. 
In spintronic devices the SOI is particularly  suitable for qubit manipulation since it can be tuned locally via electrostatic gates \cite{stepanenko04,flindt06,coish06,sanjose08,golovach10,bednarek08,fan16,gomez12,pavlowski16,pavlowski16b}. Experimentally such systems with the ability of controlling electrons
have been realised  in various semiconducting devices \cite{nadjperge12,nadjperge10,fasth05,fasth07,shin12} quantum wires.  

The simplest non-adiabatic qubit manipulation with an exact analytical solution is achieved by translating a qubit in one dimension \cite{cadez13,cadez14} in the presence of time dependent Rashba interaction \cite{nitta97,liang12}.
For quantum dots with harmonic confining potential the exact analytical solution is possible also for non-adiabatic non-Abelian Anandan phase \cite{anandan88}. However, the transformations are limited to cases of rotations with fixed axis.
Most recently limitation posed by fixed axis of spin rotation in linear systems was eliminated in a quantum ring structure \cite{kregar16,kregar16b}. 

Since exact solutions for  qubit manipulation are possible, the analysis of certain environment effects can be considered analytically \cite{lara17}: due to fluctuating electric fields, caused by the piezoelectric phonons \cite{sanjose08,sanjose06,huang13,echeveria13} or due to phonon-mediated instabilities in molecular systems with phonon assisted potential barriers, which introduce noise in the confining potentials \cite{mravlje06,mravlje08}. Electrons could be carried also by surface acoustic waves, where the noise can be caused by the electron-electron interaction \cite{giavaras06,rejec00,jefferson06}.

In this paper we concentrate on some explicit types of qubit transformation drivings, in one dimension and in a ring system. In particular,
after the introduction  we present the model in Section~2, show exact solutions of the time-dependent Schr\" odinger equation and analyse the Anandan phase. In Section~3 we analyse feasibility of arbitrary qubit transformation in a ring system. Finally, in Section 4 errors in spin-qubit transformations are analysed. Section 5 is devoted to the summary. 
\section{Anandan phase in a linear system}
\label{sec:1}
We analyse qubits  represented as spin states of an electron in a harmonic trap \cite{cadez13,cadez14}. The position of the trap $\xi(t)$ in a one-dimensional quantum wire is time dependent and controlled by the application of external electric fields. The spin is controlled by the spin-orbit interaction related to the external electric field, 
\begin{equation}\label{H}
H(t)=\frac{p^{2}}{2m^{*}}I+\frac{m^{*}\omega^{2}}{2}[x-\xi(t)]^{2}I+\alpha(t)p\,\mathbf{n}\cdot{\bm{\sigma}},
\end{equation}
where $m^{*}$ is the electron effective mass, $\omega$ is the frequency
of the harmonic trap and $\alpha(t)$ is the strength of the time dependent Rashba spin-orbit interaction.  $\bm{\sigma}$
and $I$ are Pauli spin matrices and unity operator in spin space, respectively. The spin rotation axis $\mathbf{n}$ is constant and depends 
on the crystal structure of the quasi-one-dimensional material used and 
the direction of the applied electric field \cite{nadjperge12}.  This Hamiltonian can be solved exactly \cite{cadez13,cadez14},
\begin{equation}\label{psi}
|\Psi_{ms}(t)\rangle=e^{-i\omega_{m}t}\mathcal{A}_{\alpha}(t)\mathcal{X}_{\xi}(t)|\psi_{m}(x)\rangle|\chi_{s}\rangle,
\end{equation}
\begin{equation}
\mathcal{A}_{\alpha}(t)=e^{-i[(\phi_{\alpha}(t)+m^{*}\dot{a}_{c}(t)a_{c}(t)/\omega^{2})I+\phi_A(t)\mathbf{n}\cdot{\bf\sigma}/2]}
e^{-i\dot{a}_{c}(t)p\mathbf{n}\cdot\bf{\sigma}/\omega^2}
e^{-im^{*}a_{c}(t)x\mathbf{n}\cdot\bf{\sigma}},
\end{equation}
\begin{equation}
\mathcal{X}_{\xi}(t)=e^{-i\phi_{\xi}(t)I}e^{im^{*}[x-x_{c}(t)]\dot{x}_{c}(t)I}e^{-ix_{c}(t)pI}.
\end{equation}
Here $\psi_{m}(x)$ represents the $m$-th eigenstate of a harmonic oscillator with energy $\omega_{m}=(m+1/2)\omega$ and $|\chi_{s}\rangle$ is a spinor of the electron in the eigenbasis of operator $\sigma_z$. 
The solution is determined by two unitary transformations, of spin part $\mathcal{A}_{\alpha}$ 
and charge contribution $\mathcal{X}_{\xi}$ which translate the system into the "moving frame" of both SOI
and position and transform the Hamiltonian Eq.~\ref{H} into a simple time independent hamonic oscillator Hamiltonian. 
The phase $\phi_{\xi}(t)=-\int_{0}^{t}L_{\xi}(t')\mathrm{d}t'$ is the coordinate action
integral, with $L_{\xi}(t)=m^{*}\dot{x}_{c}^{2}(t)/2-m^{*}\omega^{2}[x_{c}(t)-\xi(t)]^{2}/2$
being the Lagrange function of a driven harmonic oscillator and $x_{c}(t)$ is the
solution to the equation of motion of a classical driven oscillator
\begin{equation}\label{eq:xoscillator}
\ddot{x}_c(t)+\omega^{2}x_{c}(t)=\omega^{2}\xi(t).
\end{equation}
Another phase factor is the SOI action integral phase 
$\phi_{\alpha}(t)=-\int_{0}^{t}L_{\alpha}(t')\mathrm{d}t'$,
where $L_{\alpha}(t)=m^{*}\dot{a}_{c}^{2}(t)/(2\omega^2)-m^{*}[a_{c}(t)-\alpha(t)]^{2}/2+m^{*}\alpha^2(t)/2$
is the Lagrange function of another driven oscillator, satisfying
\begin{equation}
\ddot{a}_{c}(t)+\omega^{2}a_{c}(t)=\omega^{2}\alpha(t). \label{ac}
\end{equation}
Spin-qubits are rotated around $\mathbf{n}$ by terms proportional to operators $a_{c}(t)x$, $\dot{a}_{c}(t)p$ and the phase $\phi(t)$. We consider here only cyclic evolutions, defined by conditions  $\xi(t+T)=\xi(t)$, $\alpha(t+T)=\alpha(t)$, $x_{c}(t+T)=x_{c}(t)$, $\dot{x}_{c}(t+T)=\dot{x}_{c}(t)$, $a_{c}(t+T)=a_{c}(t)$
and $\dot{a}_{c}(t+T)=\dot{a}_{c}(t)$. The angle of the spin
rotation  is for such drivings given by the Anandan phase \cite{anandan88,cadez14},
\begin{equation}\label{phiT}
\phi=\phi(T)=-2m^{*}\int_{0}^{T}\dot{a}_{c}(t)\xi(t){\rm d}t=2m^{*}\oint_{\mathcal{C}}a_{c}[\xi]{\rm d}\xi,
\end{equation}
where $a_{c}[\xi]$ represents the contour $\mathcal{C}$ in 2D parametric space $[\xi(t),a_{c}(t)]$ for $0\leq t\leq T$. Thus
the spin rotation angle is  simply given by the area enclosed by $\mathcal{C}$.  In the limit of a very slow motion this contour will reduce to the driving curve $\alpha[\xi]$ and in this limit the area enclosed by the contour represents the Wilczek-Zee non-Abelian phase \cite{wilczek84}, {\it i.e.}, the adiabatic limit 
result $ \phi(T)\to\phi_{\rm ad}$.

\begin{figure}
\resizebox{1.\columnwidth}{!}{
\includegraphics{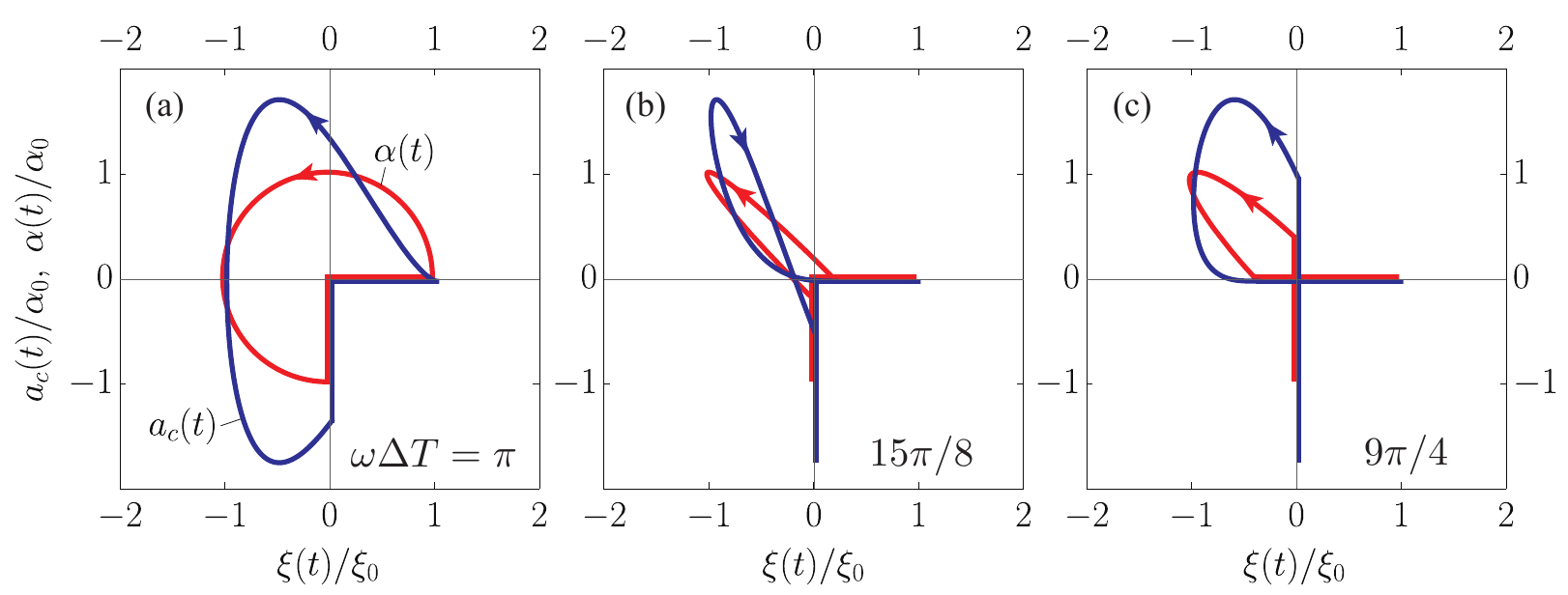}}
\caption{Contours $[\xi(t)/\xi_0,\alpha(t)/\alpha_0]$ and $[\xi(t)/\xi_0,a_c(t)/\alpha_0]$. Panels (a), (b), (c) correspond to different values of $\omega\Delta T =\pi, 15\pi/8, 9\pi/4$, respectively. Note reversed direction of motion $a_c[\xi]$ in (b) resulting in negative Anandan phase.}
\label{tiri}       
\end{figure} 

One challenging question here is: "Which, the Anandan phase $ \phi$ or the adiabatic $\phi_{\rm ad}$ phase, is for a particular driving curve larger?" A simple rule for a particular driving does not seem to be available without explicitly comparing the solutions. However, in order to elucidate this question to some extent generally we consider a family of contours of broken circular shapes represented by driving parametrized as
\begin{eqnarray}\label{broken}
\xi(t)& =& \xi_0 \sin \left( \omega t/2\right) \Theta(t)\Theta(2T_1-t),\nonumber\\
\alpha(t)& =& {\alpha_0}\xi(t-\Delta T)/{\xi_0},
\end{eqnarray}
where $\Theta(t)$ is the Heaviside step function, $T_1=2\pi/\omega$ and $\Delta T$ is the time delay. The driving is applied periodically with the cycle period $T=2T_1+\Delta T$. The responses are periodic and within one cycle given by 
\begin{eqnarray} 
x_c(t)&=&\frac{2}{3} \xi_0\left[2 \sin \left({ \omega t }/{2}\right)-\sin (\omega t)\right]\Theta(t)\Theta(2T_1-t),\nonumber\\ 
a_c(t)&=&{\alpha_0}x_c(t-\Delta T)/{\xi_0}. \nonumber
\end{eqnarray}
Various contours $[\xi(t)/\xi_0,a_c(t)/\alpha_0]\sim {\cal C}$  and $[\xi(t)/\xi_0,\alpha(t)/\alpha_0]\sim{\cal C}_{\rm ad}$  are for different $\Delta T$ presented in Fig.~\ref{tiri}.  
In the panel Fig.~\ref{tiri}(a)  note the reversion of the direction of ${\cal C}$ with respect to ${\cal C}_{\rm ad}$ which results in {\it negative} Anandan phase.
In all panels start (and end) of a cycle is at $\xi/\xi_0=1$ and $\alpha=0$ with $x_c(0)=\dot{x}_c(0)=0$ and $a_c(0)=\dot{a}_c(0)=0$.

\begin{figure}
\resizebox{1.\columnwidth}{!}{\includegraphics{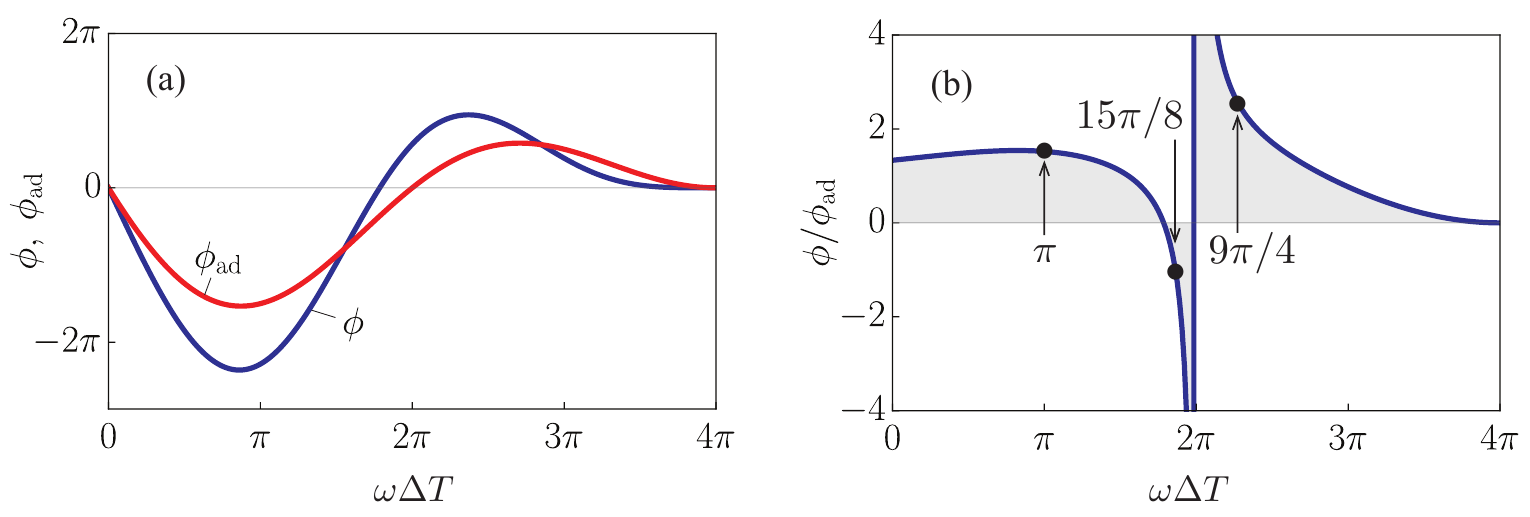}}
\caption{In panel (a) are shown the Anandan phase $\phi$ and  the Wilczek-Zee (adiabatic) phase $\phi_{\rm ad}$ plotted as a function of the delay $\omega \Delta T$. The phases are scaled by the factor $m^*\xi_0\alpha_0$. Note two points of equality where $\phi=\phi_{\rm ad}$ and that the phases change sign at different $\Delta T$. In panel (b) is presented the ratio $\phi/\phi_{\rm ad}$. Note pole at $\omega \Delta T=2\pi$, where  $\phi_{\rm ad}=0$. Arrows indicate values of $\omega \Delta T$ corresponding to special cases shown in Fig.~\ref{tiri}.  }
\label{faze}       
\end{figure}

Phases $\phi$ and $\phi_{\rm ad}$, calculated as a function of the delay $\omega\Delta T$, are presented in Fig.~\ref{faze}(a). There are two important points to be noted. (i) both curves are similar in the sense that particular phase for small $\Delta T$ is negative and by progressively larger time delay at some point changes sign and finally vanishes at $\Delta T= 2T_1$, where there is no overlap between $\xi(t)$ and $a_c(t)$. (ii) The phase curves intersect. Therefore  $\phi$ and $\phi_{\rm ad}$ can be equal for some type of driving and, moreover, the ratio $\phi/\phi_{\rm ad}$, shown in Fig.~\ref{faze}(b), can take {\it any value, positive or negative}. Since the amplitudes of drivings, $\xi_0$ and $\alpha_0$, are additional free parameters, consequently one can by changing $\Delta T$ tune the phases to any value -- independently.

\section{Arbitrary qubit transformations}

The main limitation of spin transformations, achieved by driving the electron along a straight line, is that the spin rotations are performed around a fixed axis $\mathbf{n}$. This greatly limits the range of qubit transformations that can be achieved in this manner. One way to lift this restriction is to move the electron in a two-dimensional plane, with one of the simplest motions of this kind being the motion  along a ring.

\begin{figure}
\resizebox{.9\columnwidth}{!}{\includegraphics{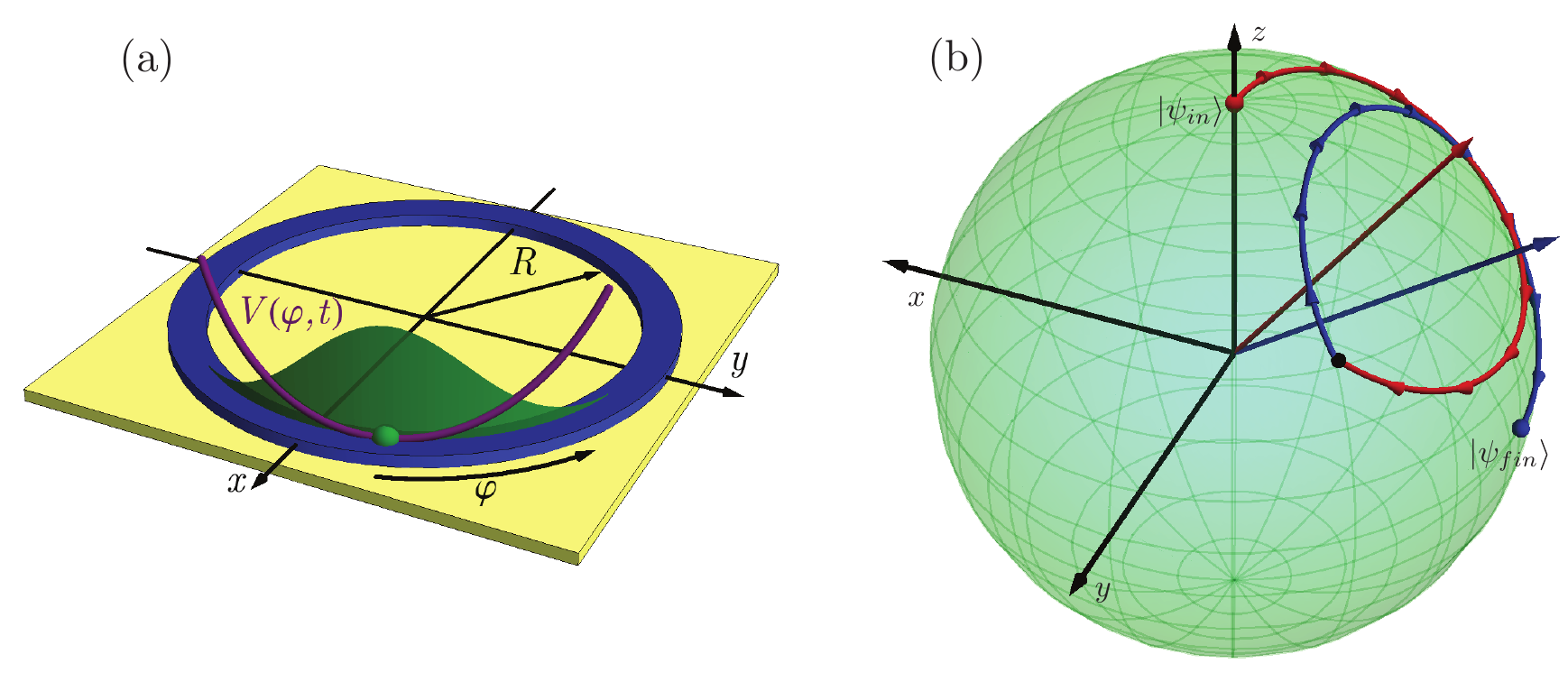}}
\caption{(a) Schematic presentation of the system. The position of the electron (green), confined by potential well $V$ (purple) on a ring  of radius $R$ (blue), is described by coordinate $\varphi$. (b) One representation of the qubit Hadamard transformation on the Bloch sphere, using the Rashba-dependent rotations $U_i^\dagger$.}
\label{fig:Had}       
\end{figure}

To describe the electron on a ring, cylindrical coordinates $r$ and $\varphi$ are a natural choice. The restriction of electron's motion to the ring is achieved by strong binding potential in radial direction, resulting in the electron occupying  the lowest radial eigenstate. Angular part of the wavefunction is then governed by an effective Hamiltonian \cite{Meijer2002}
\begin{equation} 
\label{eq:Hstart}
H = \frac{p_\varphi^2}{2 m^*} I  + \alpha(t) \left( \sigma_\rho p_\varphi - \frac{i}{2}\frac{1}{R} \sigma_\varphi \right) + V(\varphi,t) I.
\end{equation}
This Hamiltonian is effectively one-dimensional, describing the motion of the electron along the periodic coordinate $\varphi$ with its conjugate momentum $p_\varphi = -i \frac{1}{R}\frac{\partial}{\partial \varphi}$. The spin operators in the cylindrical coordinate system are given by
\begin{eqnarray}
\sigma_\rho\left( \varphi \right) &=\phantom{-} & \sigma_x \cos \varphi + \sigma_y \sin \varphi, \\
\sigma_\varphi\left( \varphi \right) &=  - &\sigma_x \sin \varphi + \sigma_y \cos \varphi.
\end{eqnarray}
Time dependent potential $V(\varphi,t)$ is a small perturbation to the potential, restricting the electron to the ring, and is used to manipulate the electron's position on a ring. As in Section~2, the motion of the electron is driven by the harmonic potential with time dependent position  as is schematically shown in Fig.~\ref{fig:Had}(a), with
\begin{equation}
V(\varphi, t) = \frac{m^{*}\omega^{2}}{2} \left[ \varphi - \xi(t) \right]^2.
\end{equation}

In order to solve the Schr\" odinger equation, we first transform the Hamiltonian with time independent transformation
\begin{equation}
\mathcal{Z}_\varphi = \exp \left(- i \frac{\varphi}{2} \sigma_z \right),
\end{equation}
which rotates the spin operators from cylindrical to Cartesian coordinates. This results in a Hamiltonian, very similar to equation (\ref{H}),
\begin{eqnarray}
H'(t) = \mathcal{Z}_\varphi^\dagger   H(t) \mathcal{Z}_\varphi 
 = \frac{ p_\varphi^{2}}{2m^{*}}I +\frac{m^{*}R^2\omega^{2}}{2}[ \varphi -\xi(t)]^{2}I + p \,\bm{\alpha}(t)\cdot{\bm{\sigma}} + \frac{1 }{8 m^* R^2}I.
\end{eqnarray}\label{H1R}

The main difference between Hamiltonians is that  the spin rotation axis in the Rashba term $\mathbf{n}$ is fixed for linear system, while here the direction of axis ${\bm \alpha}(t) =\left( \alpha(t),0,-1 \right)$
depends on the Rashba coupling and therefore changes with time. This prevents theapplication of  transformations $\mathcal{A}_{\alpha}$ and $\mathcal{X}_{\xi}$ directly to analytically calculate the time evolution of the system for an arbitrary change of parameters $\alpha(t)$ and $\xi(t)$.
However, analytical solutions can still be found for two special cases of system driving in the parameter space of $\alpha$ and $\xi$. The first is the case of the Rashba coupling being constant while the minimum of the harmonic potential is moving, and the second is adiabatic change of the Rashba coupling in static potential \cite{kregar16,kregar16b}.

To describe the time evolution of the system, we further transform the Hamiltonian with the transformation  $\mathcal{U}^{\dagger}(t)=\mathcal{A}_{\alpha}\mathcal{X}_{\xi}$ as in Section~2 but with a different form of the operator
\begin{equation}
\mathcal{A}_{\alpha} = \exp \left( -i \frac{\varphi}{2} {\bm \alpha}\cdot {\bm \sigma}  \right),
\end{equation}
resulting in a Hamiltonian of harmonic oscillator with time-dependent spin-orbit energy
\begin{eqnarray}
\label{H1R}
H''(t)= \mathcal{U} H'(t) \mathcal{U}^\dagger = \frac{ p_\varphi^{2}}{2m^{*}} +\frac{m^{*}R^2\omega^{2}}{2} \varphi^{2} + \frac{m^* \alpha(t)^2 }{2}.
\end{eqnarray}
If the Rashba coupling is constant, the time-dependent wavefunction of the system can be described in a similar manner as in linear case - a combination of eigenstates, evolving as
\begin{equation}
\label{psimst}
|\Psi_{ms}(t)\rangle=e^{-i\omega_{m}t} \mathcal{Z}_{\varphi}  \mathcal{A}_{\alpha} \mathcal{X}_{\xi(t)}|\psi_{m}(\varphi)\rangle|\chi_{s}\rangle.
\end{equation}

To describe the case of adiabatically changing Rashba coupling, it is more convenient to find a basis of Kramers states, centred at some $\xi_1$,
\begin{equation}
|\tilde{\Psi}_{ms \xi_1}(t)\rangle=e^{-i\omega_{m}t} \mathcal{Z}_{\varphi} \mathcal{X}_{\xi_1} \mathcal{A}_{\alpha(t)} |\psi_{m}(\varphi)\rangle \mathcal{Y}_{\tilde{\vartheta}_\alpha(t)}|\chi_{s}\rangle,
\end{equation}
for which the time evolution is manifested only as a change of the parameter $\alpha$ in the operator $ \mathcal{A}_{\alpha(t)}$ and
\begin{equation}
\mathcal{Y}_{\tilde{\vartheta}_\alpha(t)} = e^{-i \tilde{\vartheta}_{\alpha(t)} \sigma_y}.
\end{equation}
The rotation of spin states due to the change of the Rashba coupling $\tilde{\vartheta}_{\alpha(t)}$ can be calculated numerically and is mostly negligible in realistic systems.
If the Kramers states $|\tilde{\Psi}_{ms \xi_1}(t)\rangle$ are treated as a qubit basis, the adiabatic change of the Rashba coupling only affects the basis states, but not the coefficients of the expansion $c_s$ in the Kramers basis.
\begin{equation}
\label{eq:psit_alpha}
|\psi(\varphi,t)\rangle =  e^{i \phi_{\alpha}(t)} \sum_s  c_{s}|\tilde{\Psi}_{ms \xi_1}(t)\rangle.
\end{equation}

Driving of the electron along the ring by an external potential can also be expressed in terms of the Kramers states. If the position of the electron's wavefunction before ($\xi_{i-1}$) and after the shift of potential minimum ($\xi_{i}$) is fixed, the transformation of the wavefunction can be written as
\begin{equation}
|\psi(\varphi,t_{i-1})\rangle = \sum_s c_{i-1, s} |\tilde{\Psi}_{ms \xi_{i-1}}(t)\rangle \rightarrow |\psi(\varphi,t)\rangle = \sum_s c_{i, s}|\tilde{\Psi}_{ms \xi_{i}}(t)\rangle ,
\end{equation}
with coefficients $c_s$ transforming as $c_{i+1,s} = \sum_{s'} \chi_s^\dagger U^\dagger_{i} \chi_{s'} c_{i,s'}$.
Writing Kramers states in ordinary basis equation (\ref{psimst}) shows that the operator $U^\dagger_{i}$ is a spin rotation 
\begin{equation}
U^\dagger_i = e^{- i \frac{\gamma_i}{2} {\bf n}_{i} \cdot {\bm \sigma}}, \quad {\bf n}_{i} = (\sin  \vartheta_{\alpha_i} , 0, \cos  \vartheta_{\alpha_i}),
\end{equation}
with rotation axis ${\bf n}_{i}$ tilted by $\vartheta_{\alpha_i}$ from the $z$ to the $x$-direction and by the rotation angle $\gamma_i$, defined by
\begin{equation}
 \vartheta_{\alpha_i} = \tilde{\vartheta}_{\alpha_{i}}  - \arctan \left({2 m R \alpha_i}{} \right) , \quad \gamma_i = -\Delta \varphi_i \sqrt{1 + \left({2 m R \alpha_i}{} \right)^2},
\end{equation}
where $\alpha_i$ is the value of the Rashba coupling during the shift from positions $\xi_i$ and $\xi_{i+1}$.

If the Kramers states are considered as a qubit basis and the coefficients $c_{i s}$ parametrized as points  ${\bf r} = \left( \sin \Theta \cos \Phi, \sin \Theta \sin \Phi, \cos \Theta \right) $ on the Bloch sphere,
\begin{equation}
(c_{i, \uparrow}, c_{ i,\downarrow} ) = \left( \cos \frac{\Theta}{2}, e^{i \Phi} \sin \frac{\Theta}{2} \right),
\end{equation}
the rotation $U^\dagger_{i}$ is a simple rotation on the sphere, which gives an intuitive insight into the qubit transformations.

Here we performed a comprehensive numerical analysis of the transformation, which revealed  that any qubit transformation can be realized using the described time evolution by properly adjusting the distances by which the electron is shifted and the accompanying values of the Rashba coupling. An example of such a transformation is shown in Fig.~\ref{fig:Had}, where the Hadamard-like gate is applied to transform the qubit state  $\left| 0 \right\rangle \rightarrow \frac{1}{\sqrt{2}} \left( \left| 0 \right\rangle - \left| 1 \right\rangle   \right)$. 

\begin{figure}
\resizebox{.85\columnwidth}{!}{\includegraphics{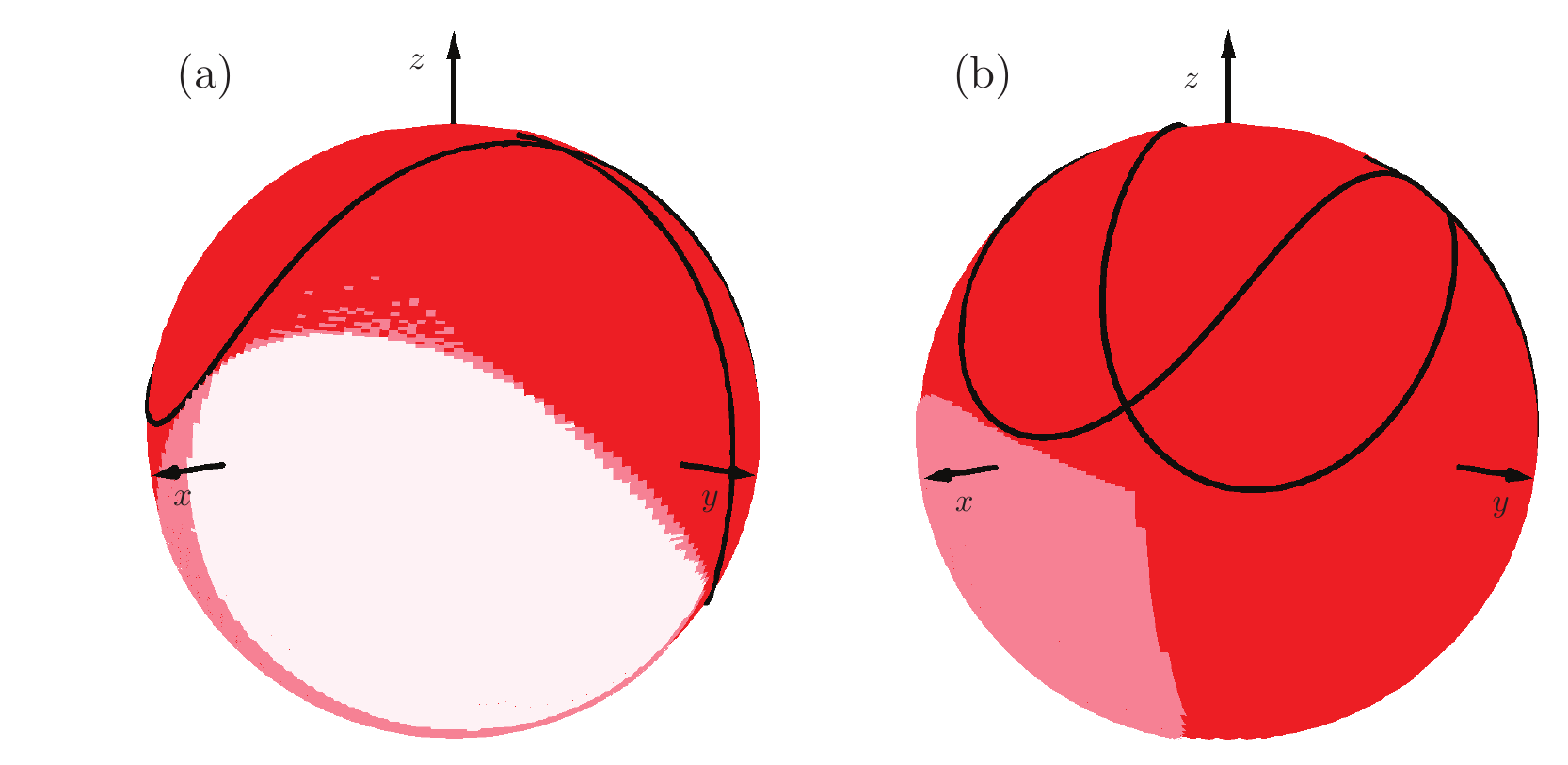}}
\caption{Figure shows parts of the Bloch sphere that are covered by one (a) or two (b) rotations of the electron around the ring for factor of amplification of the Rashba coupling $k=5$. Black line shows the area, covered with $m=2$, dark red with $m=3$ and light red with $m=4$ shifts of the electron position. For our set of $m$, the white area on (a) can only be covered for two rotations. Calculations were performed on a $200 \times 200$ grid.}
\label{fig:Bloch}       
\end{figure}

In fact, using the Monte-Carlo simulation, we demonstrated that for a sufficiently large amplification of the Rashba coupling, any qubit transformation can be achieved by properly adjusting the values of $\alpha$ during the shifts of electrons position. This is shown in Fig.~\ref{fig:Bloch}, where particular sectors of the Bloch sphere, corresponding to the qubit transformation, can be reached for one or two motions of the electron around the ring at various numbers of changes of the Rashba coupling during the revolution. Initial qubit was an eigenstate of spin along the $z$-axis (all other cases are equivalent by the symmetry of the Hamiltonian).

\section{Effect of coloured noise on qubit transformations}

For qubit transformations performed in linear systems discussed in Section~2 
the angle of spin rotation is proportional to the area in parametric space $[\xi,a_c]$ enclosed by the contour $a_c[\xi]$. In real situations there is unavoidably present some noise in driving functions $\alpha(t)$ and $\xi(t)$, {\it e.g.} due to electrostatic noise in gate potentials so it is important to analyse the stability of the qubit transformation with respect to small deviations of drivings. The change in angle of rotation is characterized by the change of contour $a_c[\xi]$. Analogue effects are present also in ring systems discussed in Section~3.
Here we show how to calculate and characterize the noise in $a_c(t)$, while the corresponding results for the position $x$ (or $\varphi$ in the case of ring systems) can easily be rederived. Once this is analyzed one can analytically predict the angle of spin rotation error since analytic results for qubit transformations are known. 

We model the noise as an additive coloured noise, $\alpha(t)=\alpha^0(t)+\delta\alpha(t)$, $\alpha^0(t)$ being the noiseless driving function and $\delta\alpha(t)$ the superimposed noise with vanishing mean $\langle\delta\alpha(t)\rangle$ and with the time autocorrelation function $\langle\delta\alpha(t')\delta\alpha(t'')\rangle=\frac{\sigma^2_\alpha}{2\tau_\alpha}e^{|t'-t''|/\tau_\alpha}$ characteristic for Ornstein-Uhlenbeck processes  \cite{uo30,wang45,masoliver92,meinrichs93}. $\sigma^2_\alpha$ is the noise intensity and $\tau_\alpha$ the correlation time.
As a general solution of equation~(\ref{ac}) $a_{c}(t)$ is given by
\begin{equation}\label{eq:harmoicsolution}
a_{c}(t)=\omega\int_{0}^{t}\sin[\omega(t-t')]\alpha(t'){\rm d}t'.
\end{equation}

\begin{figure}
\resizebox{.95\columnwidth}{!}{\includegraphics{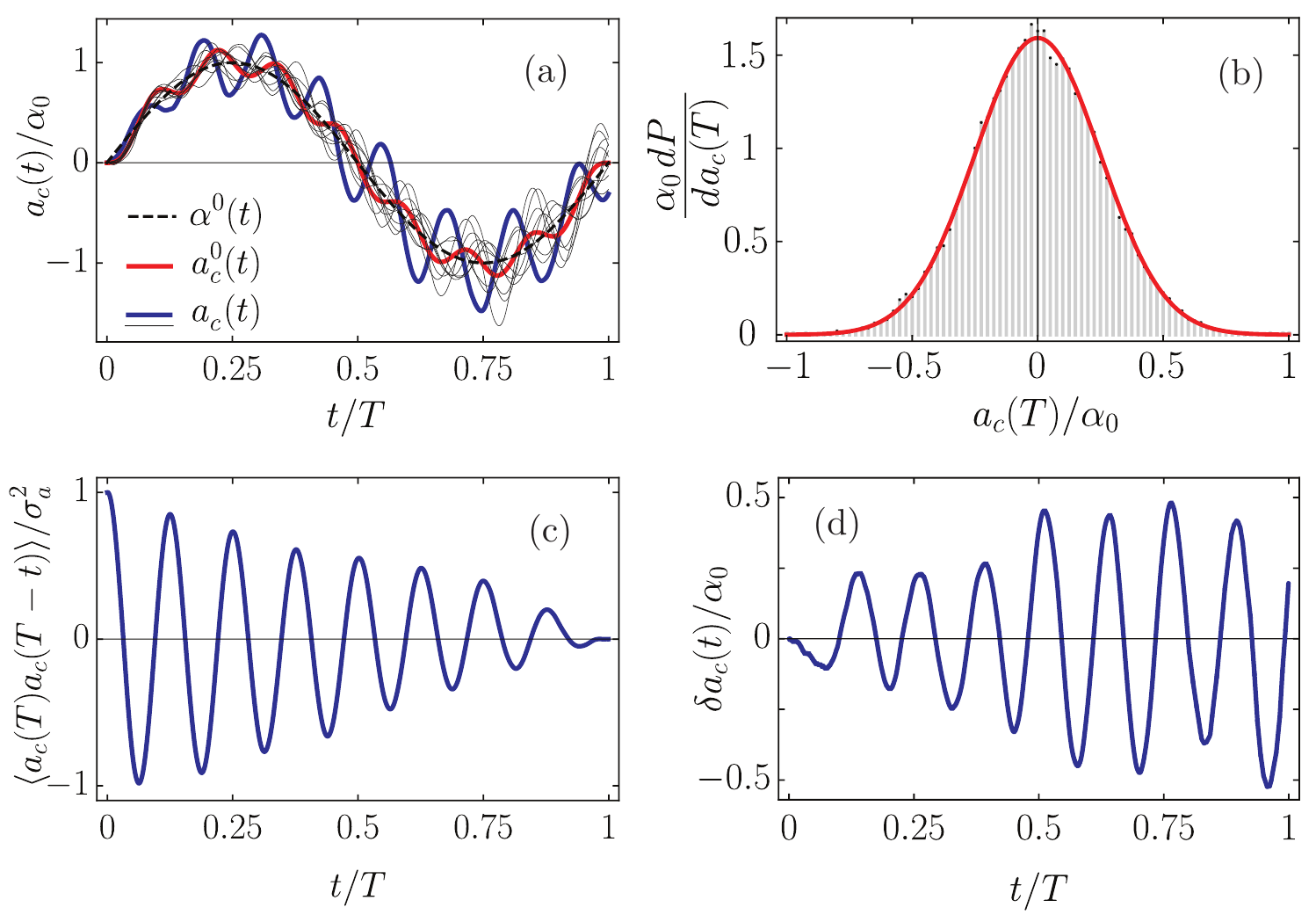}}
\caption{All figures correspond to spin-orbit response $a_c(t)/\alpha_0$ to sinusoidal driving with $n=8$ and with gaussian white noise ($\tau_\alpha\to 0$) with noise intensity $\sigma_\alpha/\alpha_0 =\frac{1}{20}\omega^{-1/2}$. In Fig.~\ref{fig:NoiseAc}(a) the dashed line marks the driving function without noise, the red one response to driving without noise and thin black lines the different realizations of response to noisy driving, one of which is marked with blue. In Fig.~\ref{fig:NoiseAc}(b) is shown the probability distribution of the final error in $a_c(T)/\alpha_0$ with red curve corresponding to analytical result and black lines to numerical one. In Fig.~\ref{fig:NoiseAc}(c) is shown the autocorrelation function of $a_c(t)$ and in (d) one realization of noise in $a_c(t)$ (note the amplitude progressively increasing with time).}
	\label{fig:NoiseAc}       
\end{figure}
In Fig.~\ref{fig:NoiseAc} an explicit example of noise in $a_c(t)$ is shown. The driving is of sinusoidal form
\begin{equation} 
\alpha^0(t)=\alpha_0 \cos(2 \pi t/T_n),
\end{equation}
with transformation times $T=T_n=   n T_1$, where $T_1=2\pi/\omega$ is the period of the confining potential and $n>1$. Driving in figures is for $n=8$. The noise added is a short correlation one ($\tau_\alpha\to 0$) which corresponds to Gaussian white noise with  $\sigma_\alpha$. Numerical calculation of $a_c(t)$ was done by summing  over discrete values,
\begin{equation}
a_c(t)=a_c^0(t) + \sum_{i=1}^{N}\alpha_ia_i,
\end{equation}
where $a_i=\omega\sin[\omega(t-t_i)]$, $t_i=i\triangle t$, $\triangle t=T/(N-1)$ and $\alpha_i$ is equal to the integral of the noise in 
a short time interval $\alpha_i=\intop_{t_i}^{t_i+\triangle t} \delta\alpha(t')\,\mathrm{d} t'$. It is a stohastic normally distributed 
value with zero mean and with variance $\sigma_{\alpha}^2\triangle t$ \cite{sed11}.
 Fig.~\ref{fig:NoiseAc}(a) shows the driving noiseless function $\alpha^0(t)$ marked with black dashed line and red curve corresponds to $a^0_c(t)$ spin-orbit response to this noiseless driving. Blue line represents response $a_c(t)$ to one realization of noisy driving $\alpha(t)$. Other responses to different noise realizations are marked with thin black lines. Final deviations of $a_c(T)$ from noiseless $a^0_c(T)$ are shown in Fig.~\ref{fig:NoiseAc}(b) as a normalized histogram (black bins) calculated from $10^7$ noise realizations. The red curve corresponds to analytic result of probability density function that is calculated in the following. As seen from the histogram, $a_c(T)$, an integral of stochastic variables, is normally distributed stochastic quantity which is in accordance with the central limit theorem. However, 
by looking at the nontrivial autocorrelation function $\langle a_c(T)a_c(T-t)\rangle$ in Fig.~\ref{fig:NoiseAc}(c) which oscillates with diminishing amplitude, the variance of distribution $\sigma_a^2$ seems to be nontrivial in time-dependence. This can be further speculated from  Fig.~\ref{fig:NoiseAc}(d) which shows bare noise in spin-orbit response $\delta a_c(t)$ as a function of time and it is evident that it oscillates with confining potential frequency and grows in amplitude.  We evaluate $\sigma_a^2$ as equal-times autocorrelation function \cite{wang45,feller71},
\begin{equation}
\sigma_a^2(t)=\omega^2\lim_{\Delta t \to0} \langle\int_{0}^{t}\sin[\omega(t-t')]\delta\alpha(t'){\rm d}t' \int_{0}^{t+\Delta t}\sin[\omega(t-t'')]\delta\alpha(t''){\rm d}t'' \rangle.
\end{equation}
This is calculated as an integral after leaving in average $\langle\rangle$ the only stochastic term $\delta\alpha(t')\delta\alpha(t'')$ and evaluating it as the time autocorrelation function. For the Ornstein-Uhlenbeck noise considered here the integrals can be evaluated exactly and the final result is that $a_c(t)$  is distributed normally with the time dependent variance
\begin{eqnarray}
\sigma_a^2(t)&=&\frac{\omega \sigma^2_\alpha}{4(1+\omega^2\tau_\alpha^2)^2}[-4\omega^2\tau_\alpha^2 e^{-t/\tau_\alpha}(\omega\tau_\alpha\cos \omega t+\sin \omega t )+\\ \nonumber
& &+2\omega t-\omega\tau_\alpha+\omega^3\tau_\alpha^2(2t+3\tau_\alpha)+(1+\omega^2\tau_\alpha^2)(\omega\tau_\alpha\cos 2\omega t-\sin 2\omega t)].\label{1d}
\end{eqnarray}
In short correlation time limit the expression simplifies to the white noise result, $\sigma_a^2(t)=\frac{1}{4}\omega\sigma_{\alpha}^{2}\left(2\omega t-{{\sin{2\omega t}}}\right)$. For large $\omega t$ the noise amplitude diverges and the reason  is that the Lorentzian noise power spectrum $ \sigma^2_\alpha/[1+(2 \pi f \tau_\alpha)^2]$ considered here consists of different driving frequencies including the resonant value $\omega=2\pi f$ which, similar to the one-dimensional random walk problem \cite{wang45}, results in the asymptotic response $\sigma_{a}^2(t) \propto t$. This indicates that fast transformations are preferable since less noise is produced. Qualitatively the same arguments are valid also for qubit transformations on ring systems considered in Section~3.

\section{Summary}

Holonomic spin manipulation in linear systems  is feasible if one can control the position of the electron $\xi$ and the strength of the  Rashba coupling $\alpha$. In the space of these two driving parameters $[\xi,\alpha]$ an arbitrary contour determines the angle of the qubit rotation in the case of adiabatic transformation. For a broad range of integrable drivings exact solutions are possible and a natural question arises: Which one of the non-adiabatic and adiabatic transformations leads to larger or smaller qubit rotation? In this paper we demonstrated that the answer crucially depends on the contour in spaces $[\xi,\alpha]$ and $[\xi,a_c]$. In particular, we showed that compared to the adiabatic result some non-adiabatic transformation angles can be larger, while for other transformations smaller. Both angles can also be equal for some contours. There seems to be no general rule.

The main shortage of qubit transformations in linear systems is the restriction to transformations represented by rotations around a fixed axis. This limitation is released if the electron can be moved on a ring system. Exact solutions of qubit dynamics are available, however, the corresponding equations do not allow to analytically determine driving parameters for arbitrary final qubit state. This was the motivation to analyse various driving schemes numerically and we demonstrated that an arbitrary final state on the Bloch sphere is reachable providing corresponding drivings.

To conclude, we examined in detail also the influence on qubit transformations due to the noise in drivings. Since analytical treatment of several driving schemes is possible one can analyse also the effects of noise exactly. We demonstrated how the errors in driving give rise to variance in the spin-orbit response function. It is shown how one can analyse the effects of a general coloured noise and as an example, we show the result for the Ornstein-Uhlenbeck noise, also in the limit of short correlation times (white noise).  Analytical results for autocorrelation function and time dependent errors are tested numerically.

\vskip 5 mm
\noindent {\bf Acknowledgements}
A.R. and L.U. acknowledge partial support from the Slovenian Research Agency under contract no. P1-0044 and T.{\v C}. the support by the National Natural Science Foundation of China (NSFC) Grant No. 11650110443.


\begin{thebibliography}{}
\bibitem{wolf01} Wolf S A, Awschalom D D, Buhrman R A, Daughton J M, von Moln\'{a}r S, Roukes S M, Chtchelkanova A Y and Treger D M 2001 \textit{Science} \textbf{294} 1488
\bibitem{hanson07} Hanson R, Kouwenhoven L P, Petta J R, Tarucha A and Vandersypen L M K 2007 \textit{Rev. Mod. Phys.} \textbf{79} 1217
\bibitem{Koopens06} Koopens F H L, Buizert C, Tielrooij K J, Vink I T, Nowack K C, Meunier T, Kouwenhoven L P and Vandersypen L M K 2006 \textit{Nature} \textbf{442} 766
\bibitem{dresselhaus55} Dresselhaus G 1955 \textit{Phys. Rev.} \textbf{100} 580
\bibitem{bychkov84}  Bychkov Y A and Rashba E I 1984 \textit{J. Phys. C: Solid State Phys.} \textbf{17} 6039
\bibitem{stepanenko04} Stepanenko D and Bonesteel N E 2004 \textit{Phys. Rev. Lett.} \textbf{93} 140501
\bibitem{flindt06} Flindt C, S\o rensen A S and Flensberg K 2006 \textit{Phys. Rev. Lett.} \textbf{97} 240501
\bibitem{coish06} Coish W A, Golovach V N, Egues J C and Loss D 2006 \textit{Phys. Status Solidi (b)} \textbf{243} 3658-72
\bibitem{sanjose08} San-Jose P, Scharfenberger B, Sch{\"o}n G, Shnirman A and Zarand G 2008 \textit{Phys. Rev. B} \textbf{77} 045305
\bibitem{golovach10} Golovach V N, Borhani M and Loss D 2010 \textit{Phys. Rev. A} \textbf{81} 022315
\bibitem{bednarek08} Bednarek S and Szafran B 2008 \textit{Phys. Rev. Lett.} \textbf{101} 216805
\bibitem{fan16} Jingtao F, Yuansen C, Gang C, Liantuan X, Suotang J and Franco N 2016 \textit{Scientific reports} \textbf{6} 38851
\bibitem{gomez12} G\' omez-Le\' on A and Platero G 2012 \textit {Phys. Rev. B} {\bf 86} 115318
\bibitem{pavlowski16}  Pawlowski J, Szumniak P and Bednarek S 2016 \textit {Phys. Rev. B}   \textbf{93} 045309  
\bibitem{pavlowski16b} Pawlowski J, Szumniak P and Bednarek S 2016 \textit{Phys. Rev. B}  \textbf{94} 155407  
\bibitem{nadjperge12} Nadj-Perge S, Pribiag V S, van den Berg J W G, Zuo K, Plissard S R, Bakkers E P A M, Frolov S M and Kouwenhoven L P 2012 \textit{Phys. Rev. Lett.} \textbf{108} 166801
\bibitem{nadjperge10} Nadj-Perge S, Frolov S M, Bakkers E P A M and Kouwenhoven L P 2010 \textit{Nature} \textbf{468} 1084-7
\bibitem{fasth05} Fasth C, Fuhrer A, Bj\"{o}rk M T and Samuelson L 2005 \textit{Nano Lett.} \textbf{5} 1487-90
\bibitem{fasth07} Fasth C, Fuhrer A, Samuelson L, Golovach V N and Loss D 2007 \textit{Phys. Rev. Lett.} \textbf{98} 266801
\bibitem{shin12} Shin S K, Huang S, Fukata N and Ishibashi K 2012 \textit{Appl. Phys. Lett.} \textbf{100} 073103
\bibitem{cadez13} \v Cade\v z T, Jefferson H J and Ram\v{s}ak A 2013 \textit{New Journal of Physics} \textbf{15} 013029
\bibitem{cadez14} \v Cade\v z T, Jefferson H J and Ram\v{s}ak A 2014 \textit{Phys. Rev. Lett.} \textbf{112} 150402
\bibitem{nitta97} Nitta J, Akazaki T, Takayanagi H and Enoki T 1997 \textit{Phys. Rev. Lett.} \textbf{78} 1335
\bibitem{liang12} Liang D and Gao X P A 2012 \textit{Nano Lett.} \textbf{12} 3263
\bibitem{anandan88} Anandan J 1988 \textit{Phys. Lett. A} \textbf{133} 171
\bibitem{kregar16} Kregar A, Jefferson J H and Ram\v{s}ak A 2016 \textit{Phys. Rev. B} \textbf{93} 075432 
\bibitem{kregar16b} Kregar A and Ram\v{s}ak A 2016 \textit{Int. J. Mod. Phys. B} \textbf{30} 1642016
\bibitem{lara17} Ul{\v c}akar L and Ram{\v s}ak A, 2017  \textit{New Journal of Physics} \textbf{19} 093015
\bibitem{sanjose06} San-Jose P, Zarand G, Shnirman A and Sch\" on G 2006 \textit{Phys. Rev. Lett.} \textbf{97} 076803
\bibitem{huang13} Huang P and Hu X 2013 \textit{Phys. Rev. B} \textbf{88} 075301
\bibitem{echeveria13} Echeverr\'{i}a-Arrondo C and Sherman E Y 2013 \textit{Phys. Rev. B} \textbf{87} 081410(R)
\bibitem{mravlje06} Mravlje J, Ram\v{s}ak A and Rejec T 2006 \textit{Phys. Rev. B} \textbf{74} 205320 
\bibitem{mravlje08} Mravlje J and Ram\v{s}ak A 2008 \textit{ Phys. Rev. B} \textbf{78} 235416 
\bibitem{giavaras06} Giavaras G, Jefferson J H, Ram\v{s}ak A, Spiller T P and Lambert C 2006
{\it Phys. Rev. B} {\bf 74} 195341
\bibitem{rejec00} Rejec T., Ram{\v s}ak A., and Jefferson J. H. 2000 \textit{J. phys., Condens. matter} \textbf{12} L233
\bibitem{jefferson06} Jefferson J H, Ram\v{s}ak A and Rejec T 2006 {\it Europhys. Lett.} {\bf 74} 764
\bibitem{wilczek84} Wilczek F and Zee A 1984 \textit{Phys. Rev. Lett.} {\bf 52} 2111
\bibitem{Meijer2002} Meijer F E, Morpurgo A F and Klapwijk T M 2002 \textit{Phys. Rev. B}
\textbf{66} 033107
\bibitem{uo30} Uhlenbeck G E and Ornstein L S 1930 \textit{Phys. Rev.} \textbf{36} 823
\bibitem{wang45} Wang M C and Uhlenbeck G E 1945 {\it Rev. Mod. Phys.} {\bf 17} 323
\bibitem{masoliver92} Masoliver J 1992 {\it Phys. Rev. A} {\bf 45} 706 
\bibitem{meinrichs93} Heinrichs J 1993 {\it Phys. Rev. E} {\bf 47} 3007
\bibitem{feller71} Feller W 1971  {\it An Introduction  to  Probability  Theory  and  its  Applications}  Vol. 1,2.  (New York, Wiley) 
\bibitem{sed11} Sokolov I M, Ebelling W and Dybiec B 2011 \textit{Phys. Rev. E} \textbf{83} 041118 

\end{thebibliography}
\end{document}